\documentclass[preprintnumbers,prl,amsmath,amssymb,superscriptaddress, twocolumn]{revtex4-1}
\usepackage{graphicx}
\usepackage{endnotes}
\usepackage{bm}
\usepackage{dcolumn}

\newcommand {\litsearch}{/Users/fmg12/Documents/papers/litsearch}
\begin{document}


\title{Unconventional superconductivity in the layered iron germanide YFe$_2$Ge$_2$}

\author{Jiasheng~Chen}
\affiliation{Cavendish Laboratory, University of Cambridge, Cambridge CB3 0HE, United Kingdom}
\author{Konstantin~Semeniuk}
\affiliation{Cavendish Laboratory, University of Cambridge, Cambridge CB3 0HE, United Kingdom}
\author{Zhuo~Feng}
\affiliation{London Centre of Nanotechnology, University College
  London, London WC1H 0AH, United Kingdom}
\author{Pascal~Reiss}
\affiliation{Cavendish Laboratory, University of Cambridge, Cambridge CB3 0HE, United Kingdom}
\author{Yang~Zou}
\affiliation{Cavendish Laboratory, University of Cambridge, Cambridge CB3 0HE, United Kingdom}
\author{Peter~W.~Logg}
\affiliation{Cavendish Laboratory, University of Cambridge, Cambridge CB3 0HE, United Kingdom}
\author{Giulio~I.~Lampronti}
\affiliation{Department of Earth Sciences, University of Cambridge, Cambridge CB2 3EQ, United Kingdom}
\author{F.~Malte~Grosche}
\email{fmg12@cam.ac.uk}
\affiliation{Cavendish Laboratory, University of Cambridge, Cambridge CB3 0HE, United Kingdom}

\date{\today}


\begin{abstract}
\noindent
The iron-based intermetallic YFe$_2$Ge$_2$ stands out among
  transition metal compounds for its high Sommerfeld coefficient of
  the order of $100~\mathrm {mJ/(mol K^2)}$, which signals strong
  electronic correlations. A new generation of high quality samples of
  YFe$_2$Ge$_2$ show superconducting transition anomalies
  below 
  $1.8~\mathrm{K}$ in thermodynamic as well as transport measurements, establishing that
  superconductivity is intrinsic in this layered iron compound outside the known
  superconducting iron pnictide or chalcogenide families. The Fermi
  surface geometry of YFe$_2$Ge$_2$ resembles that of KFe$_2$As$_2$ in
  the high pressure collapsed
  tetragonal phase, in which superconductivity 
  at temperatures as high as $10~\rm K$ has recently been reported,
  suggesting an underlying connection
  between the two systems.

\end{abstract}

\maketitle 
\noindent
Since the discovery of superconductivity in LaFePO \cite{kamihara06},
numerous iron-based superconductors have been identified within
diverse structure families, all of which combine iron with a group-V
(pnictogen) or group-VI (chalcogen) element. Unconventional
superconductivity is extremely rare among transition metal compounds
outside these layered iron systems and the cuprates, and it is almost
universally associated with highly anisotropic electronic properties
and nearly 2D Fermi surface geometries.
This contrasts with the comparatively isotropic, 3D electronic structure of the
iron germanide YFe$_2$Ge$_2$ \cite{avila04}, in which resistive and
magnetic signatures of superconductivity have recently been reported
\cite{zou14,kim15},  motivating
competing scenarios for the nature of the pairing mechanism
\cite{subedi14,singh14}.  YFe$_2$Ge$_2$
shares key properties with the alkali metal iron
arsenides (K/Rb/Cs)Fe$_2$As$_2$: it has the same ThCr$_2$Si$_2$
structure, featuring square lattice iron layers, its low temperature
heat capacity Sommerfeld coefficient
is similarly enhanced, and antiferromagnetic order can be
induced by chemical substitution \cite{ran11}.  Recent x-ray
absorption and photoemission studies have demonstrated the presence of
large fluctuating Fe-moments in YFe$_2$Ge$_2$ \cite{sirica15},
suggesting that this system is close to the border of magnetism.
There is an important difference, however: although YFe$_2$Ge$_2$
appears at first sight to be isoelectronic 
to
the alkali metal iron arsenide superconductors, the existence of Ge-Ge
bonds in YFe$_2$Ge$_2$, contrasting with the absence of As-As bonds in
the arsenides, causes the Fe oxidation state and consequently the
electronic structure to differ from that of the arsenides.

\begin{figure}
\centerline{\includegraphics[width=\columnwidth]{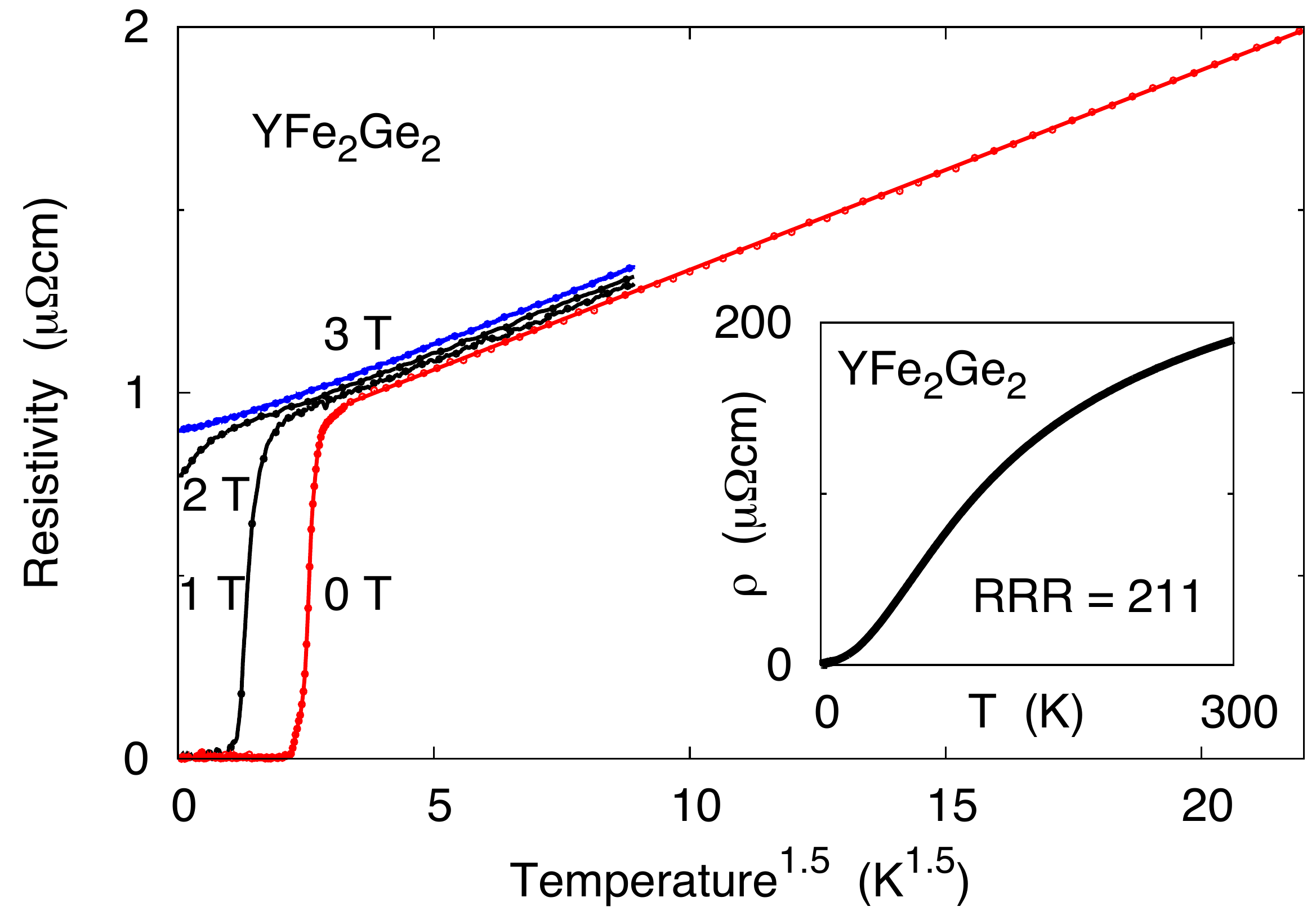}}
\caption{Electrical resistivity of YFe$_2$Ge$_2$ versus temperature, displaying a
  sharp superconducting drop of the resistivity with mid-point at
  1.83 K below a $T^{3/2}$ normal state temperature
  dependence. (inset) Temperature dependence of
  the resistivity up to room temperature. }
\label{YFGRes}
\end{figure}

Because initial experimental studies 
have failed to produce thermodynamic evidence for a bulk
superconducting transition in YFe$_2$Ge$_2$, the possibility of
filamentary superconductivity from alien phases, advanced also in
\cite{felner15}, has held back further work on this material.  Here,
we present transport and thermodynamic evidence for a bulk
superconducting transition in YFe$_2$Ge$_2$ obtained in a new
generation of high quality samples resulting from a comprehensive
programme of growth optimisation. This confirms the intrinsic nature of
superconductivity in YFe$_2$Ge$_2$ and motivates further
investigations into the nature of its unconventional superconducting
and anomalous normal state. We note, also, the striking similarity between the
electronic structure of YFe$_2$Ge$_2$ and that of KFe$_2$As$_2$ in the
pressure induced collapsed tetragonal state, which suggests that the two
systems share a common pairing mechanism.



Polycrystalline ingots of YFe$_2$Ge$_2$ were obtained by radio
frequency induction melting on a water-cooled copper boat in an argon
atmosphere.  To circumvent the formation of stable Y-Ge alloys,
YFe$_2$ was first grown from the elements (Y 3N, Fe 4N). Together with
elemental Ge (6N) this was then used to grow stoichiometric as well as
slightly off-stoichiometric YFe$_2$Ge$_2$. The melt was quenched and
then annealed in argon at $1250^\circ~\rm C$ for 1 hour, followed by
further annealing in vacuum at $800^\circ~\rm C$ for 8 days. 
 More than 20 ingots with varying nominal starting compositions
have been produced, reaching up to four times higher RRR than those reported previously \cite{zou14}.


The electrical resistance was measured using a standard four-terminal AC technique in an adiabatic demagnetisation refrigerator to $0.1\,\mathrm{K}$ and in a Quantum Design Physical Properties Measurement System (PPMS) to $<0.4 ~\mathrm{K}$. Data were scaled at 300 K to the published high temperature resistivity \cite{avila04}. The specific heat capacity was measured in a PPMS to below $0.4\,\mathrm{K}$. 
X-ray studies \cite{SuppInfo} confirm
the quality and composition of our samples. 
Our samples are typically $99\%$
phase pure, and the dominant impurity phase 
is a ferromagnetic Fe-Ge alloy with
composition approximately Fe$_{0.85}$Ge$_{0.15}$.  The
electronic structure was calculated using the Generalized Gradient
Approximation \cite{Perdew96} with Wien2k
\cite{Wien2k}. Experimentally determined lattice parameters were used
for YFe$_2$Ge$_2$ and for KFe$_2$As$_2$ at ambient pressure and
at a pressure of $21~\mathrm{GPa}$ \cite{ying15}
(Tab.~\ref{tab:cryst}). $Rk_{max}\!=\!7.5$ and 100,000 $k$-points were
used (6768 $k$-points in the irreducible Brillouin zone), and spin orbit coupling and relativistic local orbitals were included.  The fractional internal position of the Ge or As layer,
$z$, the only free internal coordinate, was optimised numerically,
resulting in $z =0.3699$ in YFe$_2$Ge$_2$ and $z= 0.3675$ in collapsed
tetragonal KFe$_2$As$_2$.

\begin{figure}
\centerline{\includegraphics[width=\columnwidth]{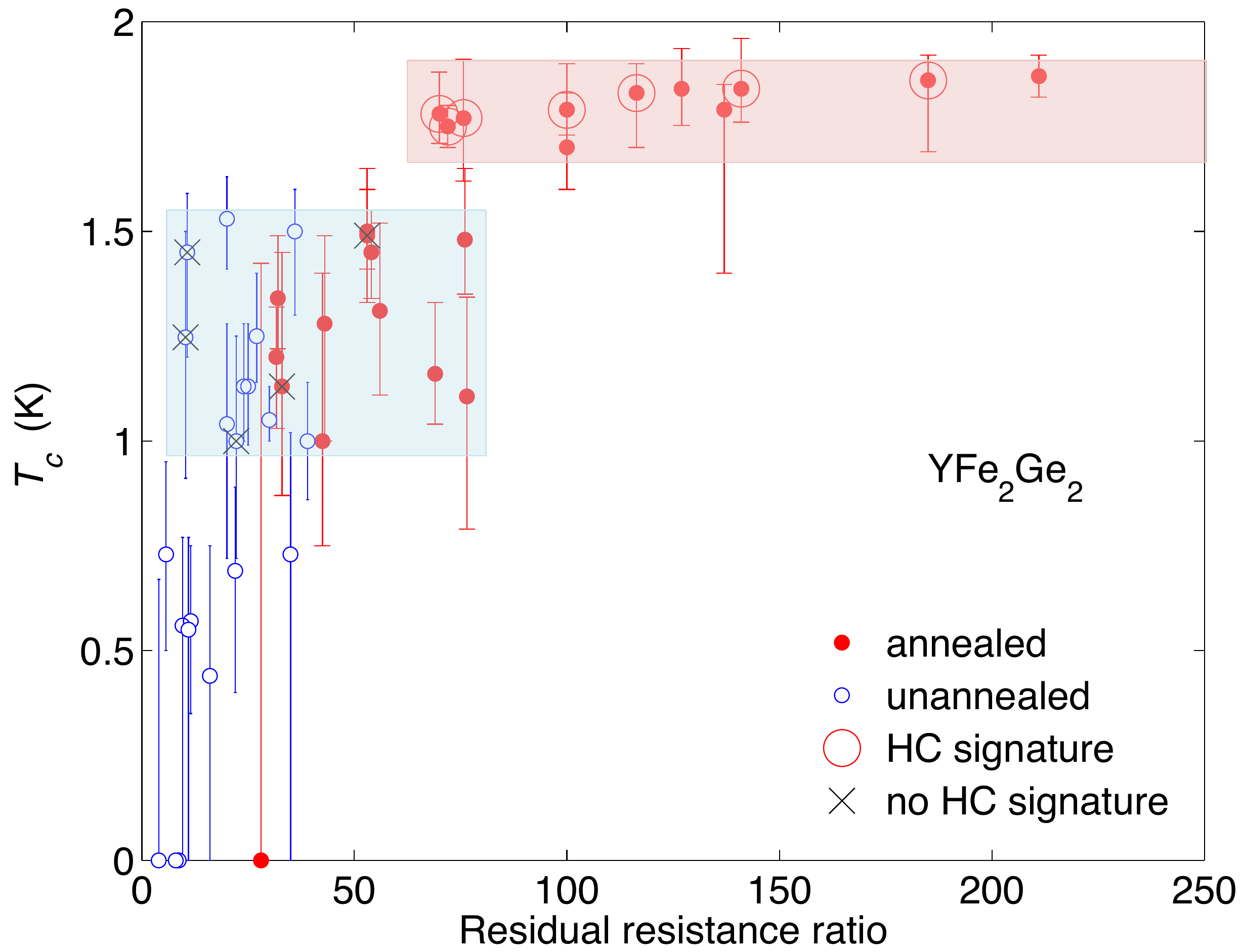}}
\caption{Correlation between superconducting transition temperatures
  and residual resistance ratio
  $\mathrm{RRR}=\rho(300\,\mathrm{K})/\rho(2\,\mathrm{K})$. Data
  points and error bars show resistive transition mid-point
  temperature and transition width, determined by an $80\%/20\%$
  criterion. The RRR = 211 sample (Fig.~\ref{YFGRes}) was too small for
  a heat capacity measurement, but the next best sample (RRR = 185,
  Fig.~\ref{YFGHC}) has been extracted closeby from the same
  ingot. Shaded areas illustrate the step in $T_c$ discussed in the
  text.}
\label{RRRTc}
\end{figure}



At temperatures below $10~\mathrm{K}$, the electrical resistivity of
all samples of YFe$_2$Ge$_2$ displays an unconventional power-law
temperature dependence of the form $\rho(T) \simeq \rho_0 + AT^{3/2}$
(Fig.~\ref{YFGRes}). This suggests Fermi liquid breakdown similar to
that observed in other transition metal compounds such as MnSi,
ZrZn$_2$ and NbFe$_2$ near the threshold of magnetic order
\cite{pfleiderer01d,takashima07,smith08,brando08} and is reminiscent
of the $T^{3/2}$ power-law temperature dependences reported in early
studies of KFe$_2$As$_2$ \cite{dong10} and CsFe$_2$As$_2$
\cite{hong13}. The dependence on residual resistivity of the resistivity exponent, which is
reported to reach the Fermi liquid value of 2 in the cleanest samples of
(K/Cs)Fe$_2$As$_2$
\cite{hardy13}, might be attributed to the hot
spot/cold spot scenario for scattering from nearly critical
antiferromagnetic fluctuations \cite{hlubina95b,rosch99}.

Although most samples show resistive superconducting transitions (Fig.~\ref{YFGRes}), the
midpoint transition temperature $T_c$ and the transition width 
depend strongly on growth conditions. The highest transition
temperatures and narrowest transitions were observed in those samples
which also have the highest residual resistance ratio, RRR (Fig.~\ref{RRRTc}).
Full resistive transitions are observed in most samples with RRR
values exceeding 20, and the value of $T_c$, which hovers around 1.3 K
up to RRR values of about 70 (blue shaded region in Fig.~\ref{RRRTc}),
steps up to around 1.8 K for $\mathrm{RRR} > 70$ (red shaded region in
Fig.~\ref{RRRTc}). This analysis of our data does not yet take into
account other underlying correlations which may affect RRR and $T_c$,
such as slight variations in nominal stoichiometry or the effect of
annealing on microscopic inhomogeneity. It does, however, suggest that
the samples most likely to display bulk superconductivity may be found
towards the high RRR end of Fig.~\ref{RRRTc}.

\begin{figure}
\centerline{\includegraphics[width=\columnwidth]{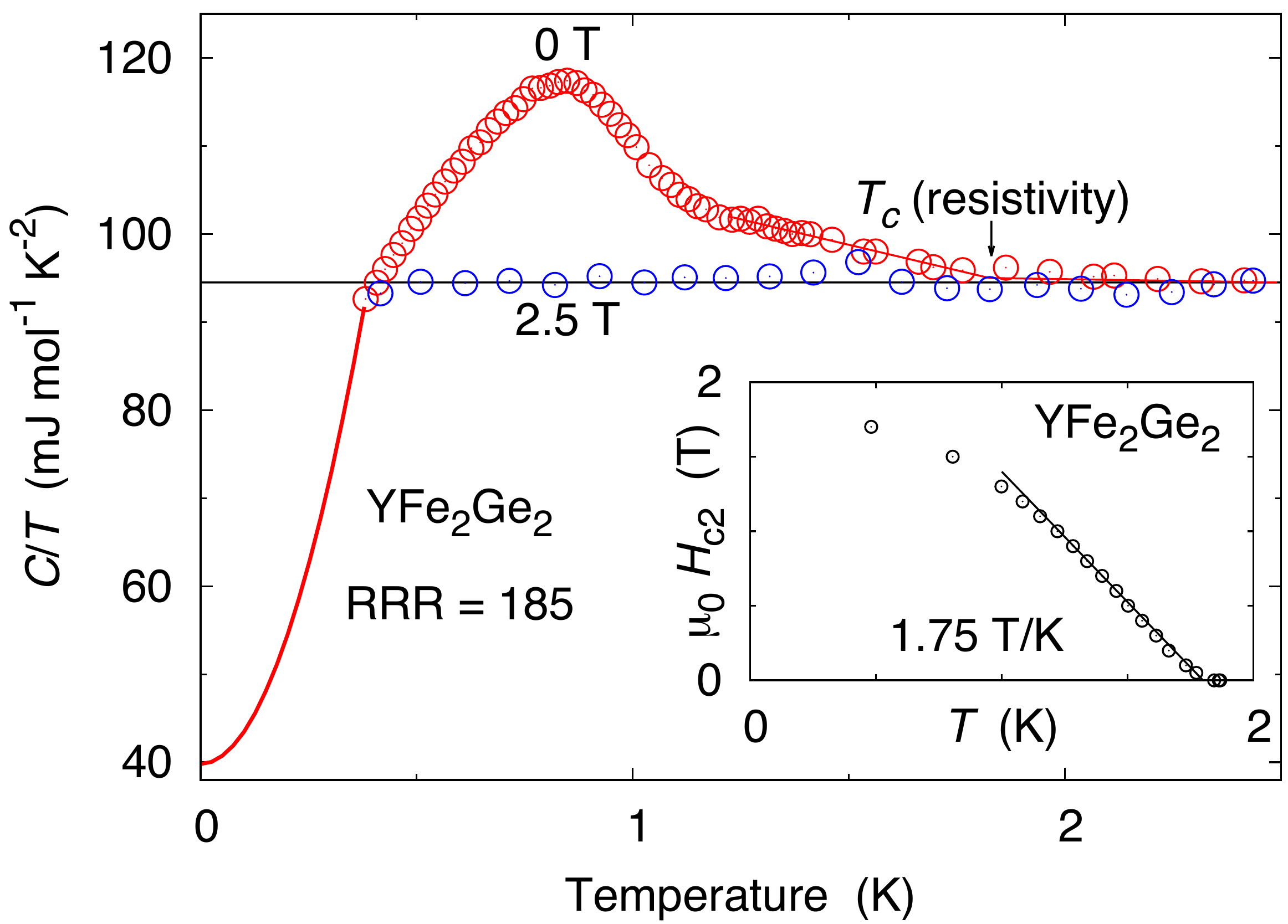}}
\caption{
  $C/T$ of YFe$_2$Ge$_2$ versus temperature. The
  extrapolation (red line) at low $T$ follows a quadratic temperature dependence
  which matches the entropy of the
  superconducting state just below $T_c$ to that of the normal state
  just above $T_c$ 
  \cite{SuppInfo}. (inset)
  Temperature dependence of the upper critical field
  determined from the mid-point of resistive
  transitions in the RRR = 211 sample. }
\label{YFGHC}
\end{figure}

Neither flux-grown samples of YFe$_2$Ge$_2$ with $\mathrm{RRR}<60$ nor
our previous generation of induction furnace-grown samples have shown
a superconducting heat capacity anomaly \cite{kim15,zou14}. By
contrast, our new generation of samples with RRR of the order of 100
or more display clear heat capacity anomalies (Fig.~\ref{YFGHC}) below
the resistive transition temperature $T_c$. The Sommerfeld ratio,
which is enhanced by an order of magnitude over the band structure value
of $\simeq 10 ~\mathrm{mJ/(mol K^2)}$ \cite{subedi14,singh14}, rises
below $T_c$, peaks at about $20\%$ above the normal state value and
then decreases rapidly below $0.8~\mathrm{K}$.

The superconducting heat capacity anomaly is suppressed in applied magnetic field, allowing a
view of the underlying normal state. The high field data show a nearly
constant Sommerfeld ratio. This suggests that the slow rise in $C/T$ shown in
\cite{zou14} may actually have been the flank of a superconducting
anomaly, broadened by sample inhomogeneity, although other
possibilities, such as closer proximity of the earlier samples to a
putative magnetic quantum critical point cannot as yet be ruled
out. Using the normal state $C/T$ measured in applied field, we can
employ an entropy-conserving equal area construction to extrapolate
the current data to lower temperature. Depending on the low
temperature form of the heat capacity, the residual Sommerfeld
coefficient in the $T=0$ limit required by balancing the entropy
reaches $24\%$, $42\%$ or $53\%$ of the normal state value,
respectively, for a linear (line-nodes), quadratic (point nodes) or
BCS-like (isotropic gap) temperature dependence of $C/T$ \cite{SuppInfo}. Similar or even larger residual $C/T$
fractions were found in early studies in the unconventional
superconductor Sr$_2$RuO$_4$ \cite{nishizaki98,*nishizaki99}, as well
as in KFe$_2$As$_2$ \cite{hardy13} and CsFe$_2$As$_2$
\cite{wang13}. SQUID magnetometry \cite{zou14} suggested
superconducting volume fractions approaching $80\%$ even in samples of
lower quality. It is likely that the present procedure still
overestimates the residual Sommerfeld ratio. If the superconducting
gap varies substantially on different sheets of the Fermi surface, as
has been proposed for KFe$_2$As$_2$ \cite{hardy14}, this can cause a
marked further downturn of the heat capacity at temperatures well
below $T_c$, not captured in the current data.

More detailed measurements to lower temperature will be necessary to
distinguish between gap scenarios, but the present data already rule
out alien phases as the origin of the superconducting heat capacity
anomaly: powder x-ray diffraction limits alien phase content to less
than $1-2\%$ even in the worst cases, of which the leading
contribution is made by ferromagnetic Fe$_{0.85}$Ge$_{0.15}$ \cite{SuppInfo}. To
obtain an apparent $50\%$ superconducting fraction from a $1\%$ alien
phase sample fraction would require the alien phase to display a
colossal normal state Sommerfeld coefficient of the order of
$5~\mathrm{ J/(mol K^2)}$, which would in turn not be consistent with
the observed critical field, the expected composition of any alien
phase and the magnitude of $T_c$.  An anomaly of this magnitude can
therefore not be consistent with the contribution of a conventional
superconducting alien phase. 

\begin{figure}

\centerline{\includegraphics[width=\columnwidth]{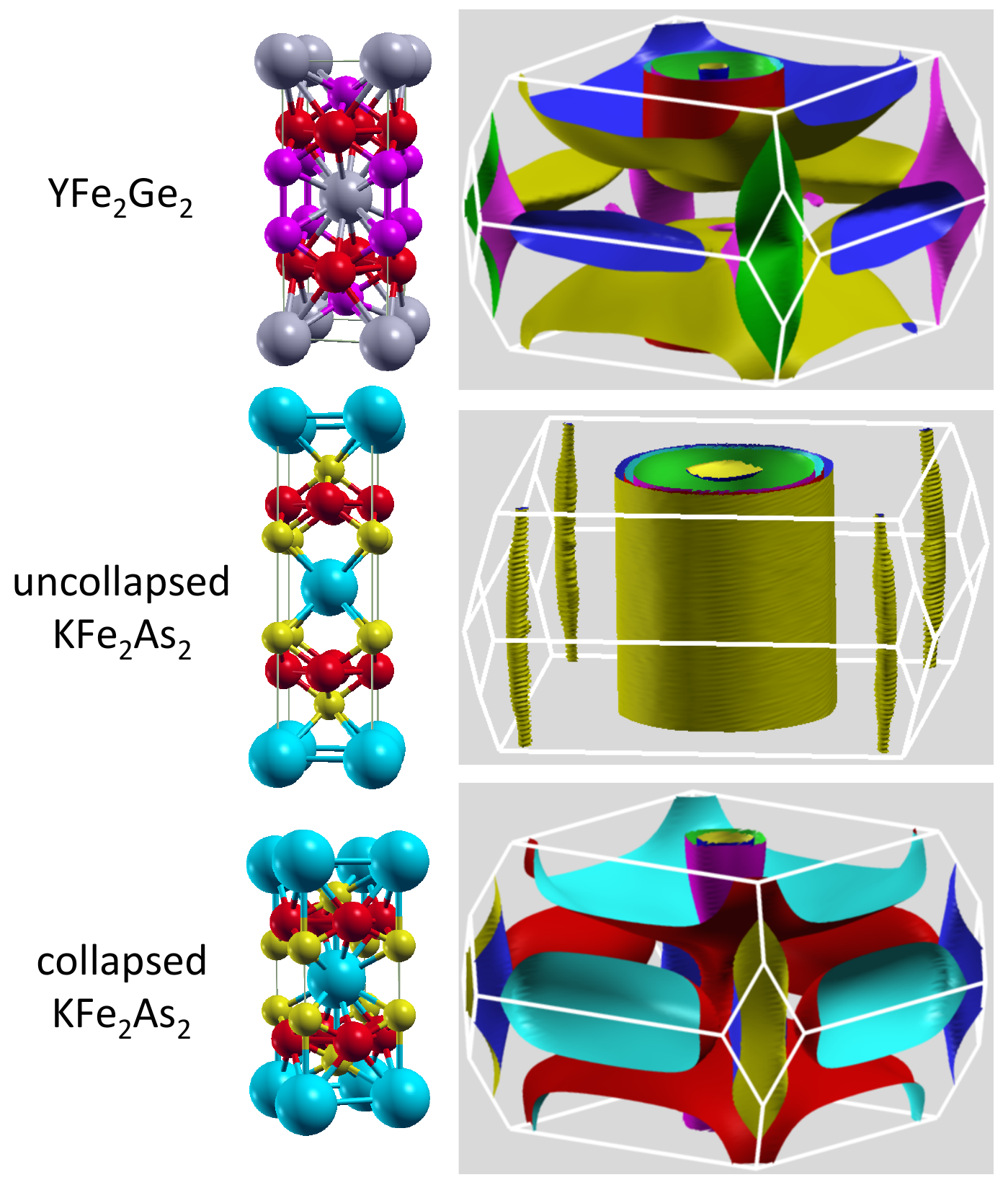}}

\caption{Fermi surface calculated within DFT for YFe$_2$Ge$_2$  and
  for KFe$_2$As$_2$ in the
  uncollapsed (uct) and collapsed (ct) tetragonal structure. The Fermi surfaces of YFe$_2$Ge$_2$ and
  uct KFe$_2$Ge$_2$ are
  fundamentally different.  Cylindrical hole sheets characterise the Fermi surface
  structure in uct KFe$_2$As$_2$, whereas 
  in  YFe$_2$Ge$_2$, nested 3D sheets around the face of the Brillouin zone (Z) and an electron
  pocket in the corner of the zone (X) are the main feature of the electronic
  structure near the Fermi energy. Conversely, the Fermi surface of ct
KFe$_2$As$_2$ is strikingly similar to that of YFe$_2$Ge$_2$. 
}
\label{YFGFS}
\end{figure}

Further information about the superconducting state can be inferred
from its response to applied magnetic field. In our new generation of
samples, the initial slope of the resistive upper critical field is
determined as $|dB_{c2}/dT| \simeq 1.75\,\mathrm{T/K}$ (inset of
Fig.~\ref{YFGHC}).  This corresponds to an extrapolated clean-limit
weak-coupling orbital-limited critical field $B^{(o)}_{c2} \simeq
0.73~ T_c ~ |dB_{c2}/dT| \simeq 2.3\,\mathrm{T}$ \cite {Helfand66},
slightly below the value reported in \cite{zou14} for a sample with a
lower $T_c$. This discrepancy may be attributed to critical field anisotropy and
preferential alignment within our polycrystals. The value for  $B^{(o)}_{c2}$ exceeds the observed critical
field in the low temperature limit, 
suggesting
that the low temperature critical field is Pauli limited. In the
standard treatment (e.g. \cite{tinkham04}), the extrapolated
orbital-limited critical field corresponds to a superconducting
coherence length $\xi_0 = \left(\Phi_0/(2\pi B^{(o)}_{c2}\right)^{1/2}
\simeq 120 ~\rm \AA$, where $\Phi_0 = h/(2e)$ is the quantum of flux.
Such a short coherence length is roughly consistent with the enhanced
quasiparticle mass and consequently low Fermi velocity indicated by
the high Sommerfeld coefficient of the specific heat capacity: we
estimate the BCS coherence length from $\xi_{BCS} = (\hbar v_F )/(\pi
\Delta)$ \cite{tinkham04,orlando79}, where $v_{F}$ is the Fermi
velocity and $\Delta$ is the superconducting gap, taken to be
$1.76 ~{\rm k_B} T_c$.  If the electronic structure of YFe$_2$Ge$_2$
(Fig.~\ref{YFGFS}) is approximated as an ellipsoidal hole
sheet around the Z point of diameter $2 ~\mathrm{\AA^{-1}}$ and height
$0.4~\mathrm{\AA^{-1}}$, its enclosed volume is $V_F \simeq
1.7~\mathrm{\AA^{-3}}$, corresponding to 1.1 carriers per formula
unit, and its surface area $S_F^{(h)} \simeq 8.2~\mathrm{\AA^{-2}}$. 
The expression for $\xi_{BCS}$ given above can be rewritten in terms
of $S_F$ as $\xi_{BCS} = V_0 S_F R/\left(1.76\cdot 12 \pi^2 \gamma_0
  T_c \right)$,
where $V_0$ is the volume per primitive unit cell, $R$ is the molar
gas constant and $\gamma_0$ is the normal state Sommerfeld
coefficient, giving $\xi_{BCS}\simeq 166~\mathrm{\AA}$, in rough
agreement with the estimate for $\xi_0$ obtained above from the
critical field measurement.


The mean free path in our samples can likewise be estimated
(e.g. \cite{orlando79}) from the Drude theory result $\ell = 6 \pi^2 h
/\left(e^2 \rho_0 S_F \right) = 15,300 \mathrm{\AA} \left(\rho_0 /
  \mathrm{\mu\Omega cm}\right)^{-1} \left(S_F /
  \mathrm{\AA^{-2}}\right)^{-1}$, where $S_F$ now includes the total
  Fermi surface area, which we estimate as $S_F \simeq 10~\mathrm
  {\AA^{-2}}$. This gives a mean free path of $\ell \simeq
  150~\mathrm{\AA}$ for samples with residual
  resistivity $\rho_0 \simeq 10~\mathrm{\mu\Omega cm}$
  ($\mathrm{RRR} \sim 20$). The observation that $T_c$ correlates with the residual resistance
ratio (Fig.~\ref{RRRTc}) and that full transitions are observed in
samples for which $\ell > \xi_0$ is consistent with
unconventional superconductivity \cite{mackenzie98}. 
Heat capacity anomalies were only observed in samples with $\rho_0 <
3~\mathrm{\mu\Omega cm}$, corresponding to $\ell > 500\mathrm
{\AA}$. We attribute this primarily to the consequences of sample
inhomogeneity: already the magnetisation measurements \cite{zou14}
showed broad transitions in lower quality samples, suggesting that the resistive $T_c$ gives an upper bound
on a distribution of transition temperatures inside the sample. This
distribution could be caused by inhomogeneity in chemical composition
or by inhomogeneity in purity, as measured by the RRR, and it would
cause the heat capacity anomaly to be smeared out in all but the best
samples. Our new heat capacity data, in which the main anomaly occurs
at a temperature of about 1K, well below the resistive transition,
shows that even in the highest quality samples there is still a
significant width to the distribution of $T_c$ values, which would in
the simplest picture follow the distribution of RRR values in the
sample.  Further complications could arise from multiband
superconductivity.



\begin{table}
\begin{center}
\begin{tabular}{|l|D{.}{.}{2.6}|D{.}{.}{3.6}|D{.}{.}{2.3}|D{.}{.}{2.3}|}
\hline
RFe$_2$X$_2$ & \multicolumn{1}{c|}{$a$} & \multicolumn{1}{c|}{$c$} & \multicolumn{1}{c|}{$c/a$} &\multicolumn{1}{c|}{X-X}\\
 & \multicolumn{1}{c|}{(\AA)} & \multicolumn{1}{c|}{(\AA)} &  & \multicolumn{1}{c|}{(\AA)}\\
\hline
YFe$_2$Ge$_2$ & 3.964(6) & 10.457(4) & 2.639 & 2.533 \\
KFe$_2$As$_2$ ($p=0$) & 3.842 & 13.861 & 3.608 & 4.089  \\
KFe$_2$As$_2$ ($21 ~\mathrm{GPa}$) & 3.854 & 9.600 & 2.491 & 2.544 \\
\hline
\end{tabular}
\end{center}
\caption{Crystallographic parameters of YFe$_2$Ge$_2$ at ambient
  pressure  \cite{zou14}, for KFe$_2$As$_2$ in the ambient pressure
  uncollapsed tetragonal phase \cite{rozsa81} and in the high pressure
  collapsed tetragonal phase \cite{ying15}. 
 Comparing  the
  Ge-Ge distance to the As-As distances illustrates
  the bond formation which accompanies the transition into the
  collapsed tetragonal phase. 
}
\label{tab:cryst}
\end{table}

The present thermodynamic evidence establishes superconductivity in
YFe$_2$Ge$_2$ as an intrinsic bulk phenomenon, motivating a more
careful look at the likely pairing mechanism. Two theoretical studies
\cite{subedi14,singh14} investigate the electronic structure of
YFe$_2$Ge$_2$, its magnetic properties and the role these could play in
determining its superconducting gap structure. Both studies arrive at
a Fermi surface structure similar to that shown in
Fig.~\ref{YFGFS}. The Fermi surface is dominated by a large
disk-shaped hole pocket enclosing the Z-point of the body-centred
tetragonal Brillouin zone, as well as a cylindrical electron pocket in
the corner of the zone. There are also several smaller hole pockets
around the Z-point. 

The calculated Fermi surface in YFe$_2$Ge$_2$ is very similar to that expected
for KFe$_2$As$_2$ in the pressure induced collapsed tetragonal phase
(Fig.~\ref{YFGFS}) \cite{nakajima15a,guterding15a}. This strongly
suggests that YFe$_2$Ge$_2$ is an isoelectronic and isostructural
reference compound to collapsed-tetragonal KFe$_2$As$_2$. The lattice
collapse in 1-2-2 arsenides is linked to the formation of
As-As bonds \cite{hoffmann85} and therefore is expected to have profound consequences
for the electronic structure, changing the Fe oxidation state to that
of YFe$_2$Ge$_2$, which features Ge-Ge bonds already at ambient
pressure (Tab. \ref{tab:cryst}). In view of
the recent surprising discovery of superconductivity at enhanced transition
temperatures exceeding $10~\mathrm{K}$ in KFe$_2$As$_2$ within the collapsed-tetragonal
phase \cite{ying15,nakajima15a}, the scenarios for superconductivity in
YFe$_2$Ge$_2$ put forward by Subedi
\cite{subedi14} and Singh \cite{singh14} assume a wider
relevance. Whereas the former argues that
the presence of an electron pocket at the zone corner and hole pockets
near the zone centre favour antiferromagnetic spin fluctuations and an $s_\pm$
order parameter wave function, the latter puts forward a 
more radical proposal: noting that magnetism with ordering wavevector
$(0, 0, 1/2)$ can be induced in
YFe$_2$Ge$_2$ by alloying with isoelectronic Lu \cite{fujiwara07,ran11}, and that this ordered
state also represents the lowest energy spin state within DFT
calculations, ferromagnetic (within the plane)
correlations could induce a triplet superconducting state. The $s_\pm$
scenario resembles the proposal which has been advanced for high
pressure KFe$_2$As$_2$ \cite{guterding15a}.

Our experimental results demonstrate that YFe$_2$Ge$_2$ undergoes a
superconducting instability at $T_c \simeq 1.8~\mathrm{K}$ out of a
strongly correlated normal state with a high Sommerfeld ratio $\gamma \sim
100~\mathrm{mJ/(molK^2)}$ and a non-Fermi liquid form for the
temperature dependence of the resistivity, $\rho(T) \simeq \rho_0 + A
T^{3/2}$. Together with 
the strong sensitivity to disorder of the resistive $T_c$ and of the
heat capacity anomaly this suggests an unconventional pairing mechanism.
Unconventional superconductivity is rare among transition metal
compounds, and YFe$_2$Ge$_2$ stands out for its comparatively
isotropic, 3D Fermi surface, when compared to the cuprates,
iron pnictides and chalcogenides, or Sr$_2$RuO$_4$.
The electronic structure of YFe$_2$Ge$_2$ resembles that of
KFe$_2$As$_2$ in the collapsed tetragonal
phase, which can be induced by applied hydrostatic pressure and in
which superconductivity with transition temperatures of the order of
$10~\mathrm{K}$ have been reported \cite{ying15,nakajima15a}.  This
establishes YFe$_2$Ge$_2$ as a reference compound for investigating
the origin of superconductivity in the collapsed tetragonal phase in
alkali metal iron arsenides, which is otherwise only accessible at
very high applied pressures.


\section{Acknowledgments}
\begin{acknowledgments}
  We thank, in particular, S. Friedemann, M. Gamza, C. Geibel,
  P. Niklowitz and G. Lonzarich for helpful discussions, and J. Dann
  and P. Brown for assistance in aspects of sample preparation and low
  temperature measurement. The work was supported by the EPSRC of the
  UK and by Trinity College.

\end{acknowledgments}

\bibliography{\litsearch/YFGrefs}

\onecolumngrid
\begin{center}
{\bf \large{Superconductivity in the layered iron germanide YFe$_2$Ge$_2$\\
 Supplemental Material}}\\
\vspace{2em}
\end{center}
\twocolumngrid

\section{Analysis of heat capacity measurements}
\begin{figure}
\centerline{\includegraphics[width=\columnwidth]{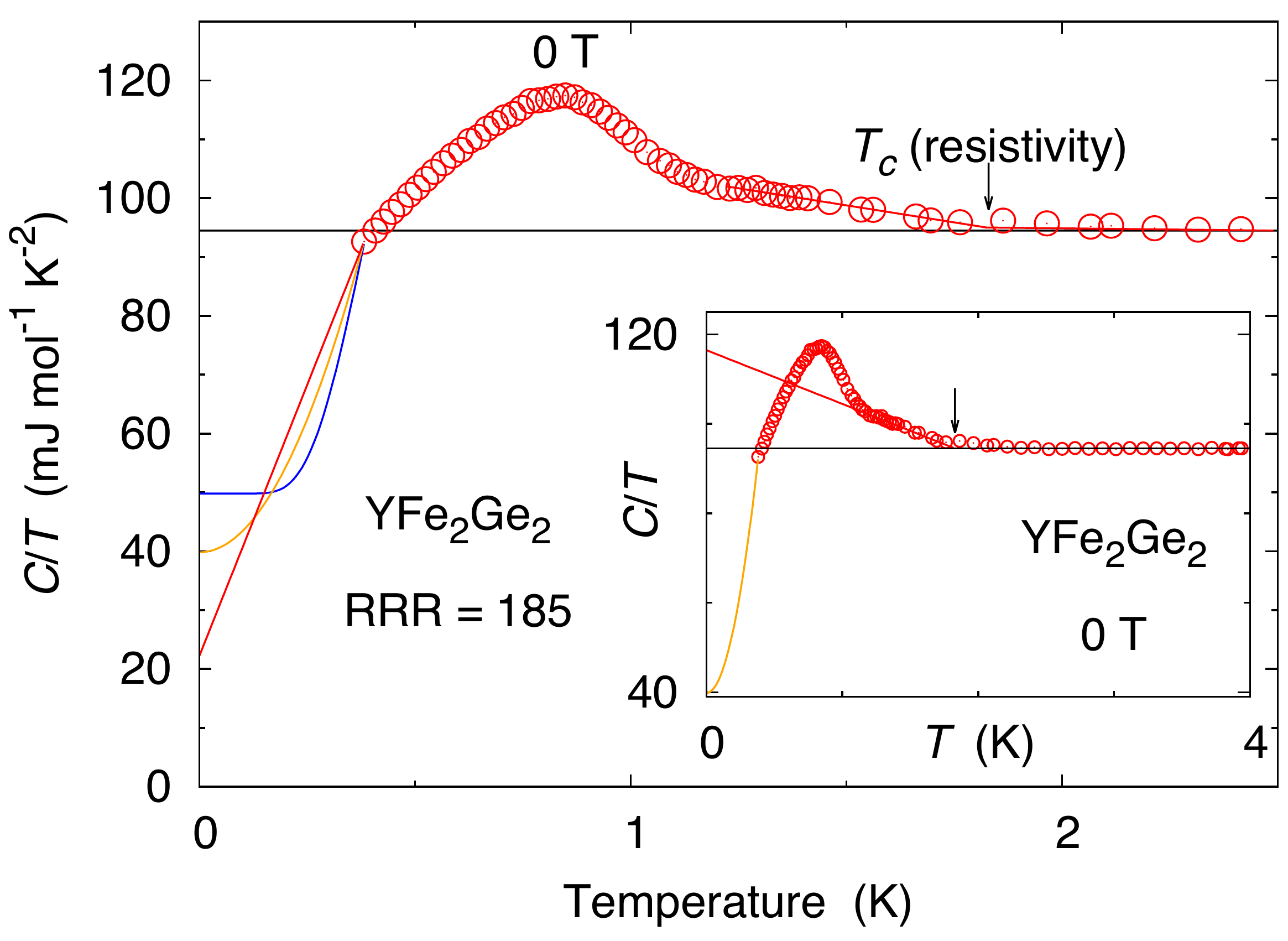}}
\caption{Illustration of extrapolation schemes for the low temperature heat capacity in YFe$_2$Ge$_2$. The
 data is the same as that shown in the manuscript. We compare a linear temperature dependence of $C/T$ (red line), a quadratic  $T$-dependence (orange line) and the $T$-dependence expected from BCS theory (blue line). The entropy of normal and superconducting states are matched at $T_c$, and the extrapolation function $C_0$ matches the measured $C$  at the lowest measured temperature. The normal state $C/T$ is taken from the measurement in high field shown in the manuscript, for which $C/T \simeq 94.5~\mathrm{mJ mol^{-1} K^{-2}}$. (inset) $C/T$ for the same sample as in the main figure, plotted over a wider $T$-range. The figure illustrates the distinct change in the slope of $C/T$ vs. $T$ near the resistive $T_c$. }
\label{YFGHCExtrapolate}
\end{figure}

\noindent
In order to extract estimates of the low temperature residual $C/T$, and thereby of the non-superconducting fraction of the sample, we have extrapolated the $C/T$ data subject to an entropy balancing constraint commonly used in this situation: the normal state entropy just above $T_c$ has to match the entropy of the superconducting state just below $T_c$. To obtain the normal state entropy, we integrate up the heat capacity in applied magnetic fields larger than the upper critical field, 
\[ 
S_n(T_c)=\int_0^{T_c} \frac{ C(H>H_{c2}, T)}{T} dT~ ,
\]
which can be approximated as $S_n (T_c) = \gamma_0 T_c$. Here, $\gamma_0$ is the Sommerfeld coefficient in the field-induced normal state, which is taken as constant (Fig.~\ref{YFGHCExtrapolate}). 

The superconducting state entropy has a known contribution within the measured temperature range, 
\[
S_1 = \int_{T_0}^{T_c} \frac{C(H=0, T)}{T} dT ~, 
\] 
where $T_0$ is the lowest temperature measured. This integral is calculated from the measured data using the trapezoidal method. A further contribution, $S_0$, then results from the extrapolation of the heat capacity data to lower temperature. It has to satisfy $S_0 + S_1 = S_n$. Moreover, we constrain the extrapolation to join the measured data point at the lowest measured temperature $T_0$: $C_0(T_0) = C(H=0, T_0)$, where $C_0(T)$ is the extrapolated heat capacity.

For an initial analysis, we have chosen three forms for the temperature dependence of $C_0(T)/T$: 

\begin{enumerate}
\item Linear, corresponding to line nodes in the gap function: 
\[
C_0(T)/T = \alpha_1 \gamma_0 + (1-\alpha_1) \beta_1 T 
\]

\item Quadratic, corresponding to point nodes in the gap function:
\[
C_0(T)/T = \alpha_2 \gamma_0 + (1-\alpha_2) \beta_2 T^2 
\]

\item The BCS form for an isotropic gap $\Delta$, which approaches $\Delta = 1.76 k_B T_c^{HC}$ in the low temperature limit, where we retain the freedom to fix a $T_c^{HC}$ different from the resistive $T_c$: 
\[ 
C_0(T)/T = \alpha_3 \gamma_0 + (1-\alpha_3) C_{BCS}(T, T_c^{HC})/T 
\] 

\end{enumerate}

Here, $\alpha_1$, $\alpha_2$ and $\alpha_3$ denote the non-superconducting fractions of the sample in the three cases. The constraints mentioned above, namely (i) matching of normal state entropy and superconducting state entropy at $T_c$, and (ii) matching of extrapolation function heat capacity and measured heat capacity at the lowest measured temperature $T_0$, make it possible to fix $\alpha$ and $\beta$ in the forms (1) and (2), above, or $\alpha$ and $T_c^{HC}$ in the BCS form (3).

A comparison of the three extrapolation schemes is shown in Fig.~\ref{YFGHCExtrapolate}. It demonstrates that independently of the details of the extrapolation scheme, the superconducting fraction is at least of order $50\%$ of the sample: (i) for the linear extrapolation of $C/T$, $\alpha_1 = 0.24$; (ii) for the quadratic extrapolation, $\alpha_2=0.42$ and (iii) for the BCS form: $\alpha_3=0.53$. The BCS extrapolation required setting a superconducting transition temperature of $T_c^{HC} = 0.72~\mathrm{K}$, however, which is well below the peak in the heat capacity plot, casting doubt on its applicability.

The inset of Fig.~\ref{YFGHCExtrapolate} shows $C/T$ over a wider temperature range, illustrating the change in slope in $C/T$ vs. $T$ near the resistive $T_c$. Whereas $C/T$ is nearly constant above $T_c$, it rises slowly below $T_c$ (orange line on the inset in Fig.~\ref{YFGHCExtrapolate}), before the main heat capacity anomaly is reached.  The rise on cooling indicated by the orange line can be attributed to a significant fraction of the sample undergoing the superconducting transition before the main part of the sample, as would be expected from the discrepancy between the resistive $T_c$ and the temperature at which the main heat capacity anomaly takes place. 

\begin{figure}
\centerline{\includegraphics[width=\columnwidth]{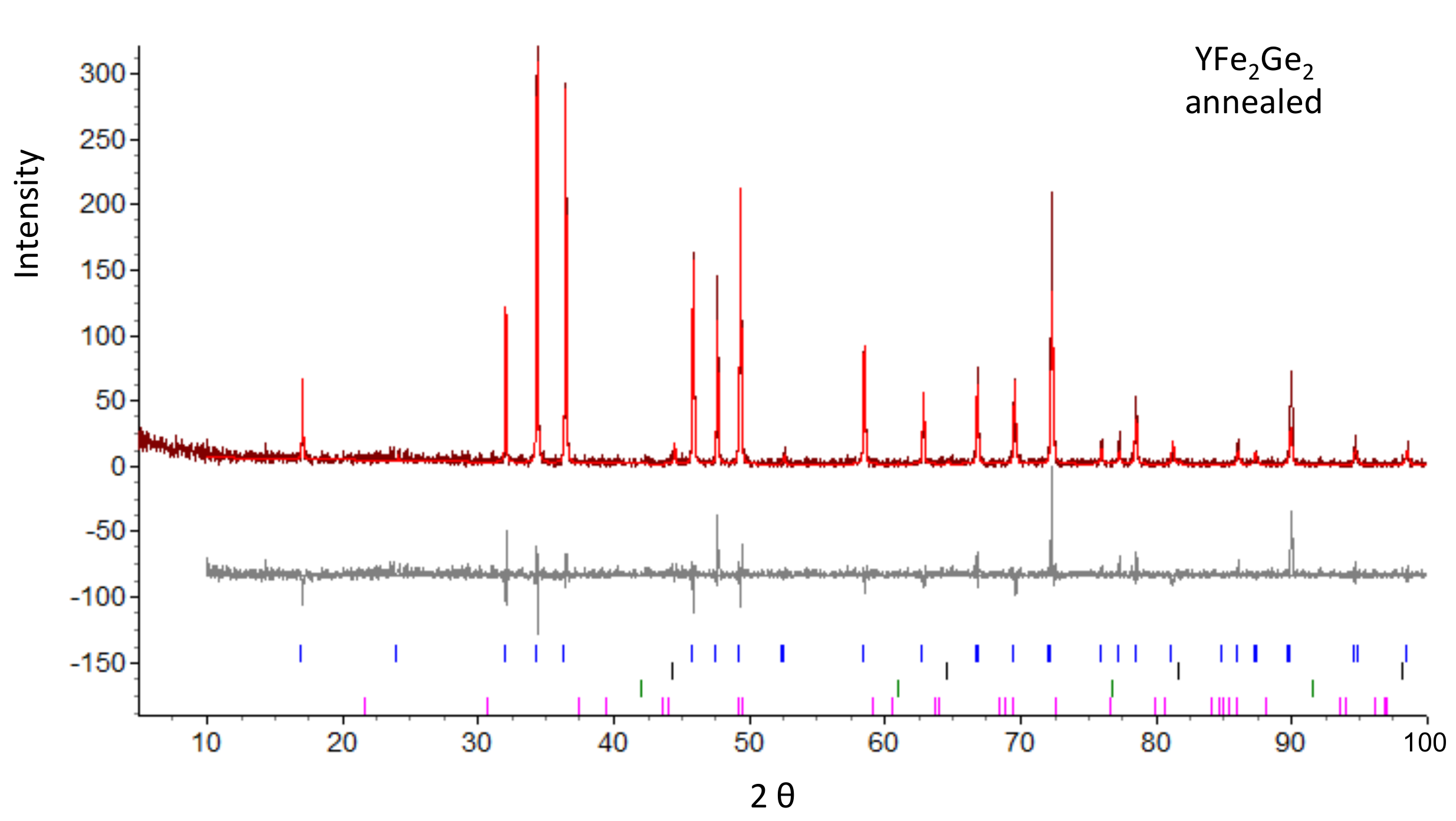}}
\caption{X-ray pattern obtained on powder from the same annealed ingot from which the heat capacity and transport samples shown in the main part of the manuscript have been extracted. Rietveld refinement yields: (i) YFe$_2$Ge$_2$: $98.73 \%$ (blue markers), (ii) Fe$_{0.85}$Ge$_{0.15}$, a bcc iron-germanium alloy with Im-3m structure: $1.16 \%$ (black markers), (iii) iron: $0.11 \%$ (green markers), and (iv) traces of FeGe$_2$ (I4mcm) at less than $0.01 \%$ (purple markers). }
\label{PowderXray}
\end{figure}

\section{Sample characterisation by powder x-ray diffraction}
\noindent 
All data were collected in Bragg-Brentano geometry on a D8 Bruker diffractometer equipped with a primary Ge monochromator for Cu K$_{\alpha 1}$ and a Sol-X solid state detector to reduce the effects of Fe fluorescence (Fig.~\ref{PowderXray}). Collection conditions were: $5-100 ^\circ$ in $2 \theta$, $0.03^\circ$ step size, 10 seconds/step, divergence slits 0.1 mm, receiving slit 0.2 mm, sample spinning. Rietveld refinements were performed with the software Topas 4.1. 

Crystal structures of all phases were retrieved from the inorganic crystal structure database: YFe$_2$Ge$_2$ (I4/mmm, ICSD reference code: 81745); Fe$_{0.85}$Ge$_{0.15}$ (Im-3m, 103493); Fe (Im-3m, 64795); FeGe2 (I4/mcm; 42519). A spherical harmonic model was applied to correct for preferred orientation of YFe$_2$Ge$_2$ within the powder. No structural parameter was refined when resolving the phase content. A shifted Chebyshev function with six parameters was used to fit the background. Peak shapes of all phases were modelled using Pseudo-Voigt functions.

\end{document}